\documentclass[12pt]{article}
\usepackage{epsf}

\newcommand\epjc[3]{Eur.\ Phys.\ J.\ C {\bf #1}, #3 (#2)}
\newcommand\ijmpa[3]{Int.\ J.\ Mod.\ Phys.\ A {\bf #1} (#2) #3}
\newcommand\npb[3]{Nucl.\ Phys.\ B {\bf #1} (#2) #3}
\newcommand\plb[3]{Phys.\ Lett.\ B {\bf #1} (#2) #3}
\newcommand\prd[3]{Phys.\ Rev.\ D {\bf #1} (#2) #3}

\newcommand{\hepph}[1]{{\tt hep-ph/#1}}

\begin{document}

\begin{titlepage}

\begin{flushright}
CLNS~01/1733\\
{\tt hep-ph/0104280}\\[0.2cm]
26 April 2001
\end{flushright}

\vspace{1.5cm}
\begin{center}
\Large\bf\boldmath
Note on the Extraction of $|V_{ub}|$ using\\
Radiative $B$ Decays
\unboldmath
\end{center}

\vspace{1.0cm}
\begin{center}
Matthias Neubert\\[0.1cm]
{\sl Newman Laboratory of Nuclear Studies, Cornell University\\
Ithaca, NY 14853, USA}
\end{center}

\vspace{0.8cm}
\begin{abstract}
\vspace{0.2cm}\noindent
The element $|V_{ub}|$ of the quark mixing matrix can be extracted with 
small theoretical uncertainties by combining weighted integrals over 
the endpoint regions of the lepton spectrum in $B\to X_u\,l\,\nu$ decays 
and the photon spectrum in $B\to X_s\gamma$ decays. The perturbative
corrections to this determination are computed at next-to-leading order
including operator mixing, which has an important impact. The effect of 
Sudakov resummation is shown to be numerically insignificant.
\end{abstract}

\vspace{2.0cm}
\centerline{\sl (Submitted to Physics Letters B)}
\vfill

\end{titlepage}

The extraction of the Cabibbo--Kobayashi--Maskawa matrix element 
$|V_{ub}|$ from charmless semileptonic $B$ decays is complicated by the
fact that, over most of phase space, there is a large background from 
decays into final states containing a charm hadron. Tight experimental 
cuts (e.g., on the charged-lepton energy or the hadronic invariant 
mass) must be applied to isolate the signal from $b\to u$ transitions. 
Accounting for such cuts theoretically is difficult, because inclusive 
decay spectra close to the kinematic endpoint are more susceptible to 
nonperturbative strong-interaction effects than fully inclusive decay 
rates. To describe the decay spectra near the boundary of phase space, 
the operator product expansion for inclusive $B$ decays must be replaced 
by a twist expansion \cite{shape,me,Fermi,Korc}. At leading power in 
$\Lambda_{\rm QCD}/m_b$, bound-state effects in the $B$ meson are 
incorporated by a shape function accounting for the ``Fermi motion'' of 
the $b$ quark. The presence of this function introduces additional 
hadronic uncertainties.

It was suggested long ago that $|V_{ub}|$ could be extracted with small 
theoretical uncertainties by combining weighted integrals over the 
endpoint regions of the lepton spectrum in $B\to X_u\,l\,\nu$ decays and 
the photon spectrum in $B\to X_s\gamma$ decays \cite{me}. The underlying 
idea is that the soft QCD interactions affecting these two spectra are 
the same and can be canceled by taking an appropriate ratio of weighted 
integrals. The result is
\begin{equation}\label{nice}
   \left| \frac{V_{ub}}{V_{cb}} \right|^2 \simeq  
   \left| \frac{V_{ub}}{V_{tb} V_{ts}^*} \right|^2 
   = \frac{3\alpha}{\pi}\,K_{\rm pert}\,
   \frac{\widehat\Gamma_u(E_0)}{\widehat\Gamma_s(E_0)}
   + O(\Lambda_{\rm QCD}/M_B) \,,
\end{equation}
where $\alpha=1/137.036$ is the fine-structure constant, and the
quantities
\begin{eqnarray}\label{rates}
   \widehat\Gamma_u(E_0) &=& \int\limits_{E_0}^{M_B/2}\!dE_l\,
    \frac{d\Gamma(B\to X_u\,l\,\nu)}{dE_l} \,, \nonumber\\
   \widehat\Gamma_s(E_0) &=& \frac{2}{M_B} 
    \int\limits_{E_0}^{M_B/2}\!dE_\gamma\,
    w(E_\gamma,E_0)\,\frac{d\Gamma(B\to X_s\gamma)}{dE_\gamma}
\end{eqnarray}
are to be determined from experiment. At tree-level, the weight function
appearing in the definition of $\widehat\Gamma_s(E_0)$ is a straight
line, $w(E_\gamma,E_0)=E_\gamma-E_0$ \cite{me}. It was shown in this 
paper that at order $\alpha_s$ there are perturbative corrections 
resulting in a matching constant $K_{\rm pert}$ at the scale 
$\mu\sim m_b$ and a logarithm, which was conjectured to result from 
renormalization-group evolution between the scales $\mu\sim m_b$ and 
$\mu\sim(\Lambda_{\rm QCD}\,m_b)^{1/2}$. The precise nature of this 
logarithm was clarified in \cite{Korc}, where the factorization of hard, 
collinear and soft contributions to the decay rate was proven to all 
orders of perturbation theory, and Sudakov logarithms were summed using
the renormalization group. In \cite{Ira}, this resummation was 
implemented in momentum (rather than moment) space.

Recently, the relation (\ref{nice}) has been evaluated for the first 
time using experimental data \cite{Ed}. It is therefore timely to update 
the calculation of the QCD corrections. In this note, we report the 
complete next-to-leading order (NLO) expression for $K_{\rm pert}$ 
including the effects of operator mixing. This is crucial for obtaining 
a renormalization-group invariant (i.e., scale and scheme-independent) 
answer. We study the residual renormalization-scale dependence of the 
result and its sensitivity to the charm-quark mass, account for power 
corrections of order $(\Lambda_{\rm QCD}/m_c)^2$, and comment on the 
numerical significance of the summation of Sudakov logarithms.

Perturbative corrections to the endpoint region of the charged-lepton 
energy spectrum in $B\to X_u\,l\,\nu$ decays were computed long ago 
\cite{Jeza}. On the other hand, the complete NLO result for the photon
energy spectrum in $B\to X_s\gamma$ decays was obtained only recently 
by combining the findings of several authors \cite{Ali,GHW,CMM,Alex}. 
Using these results, we have evaluated the corrections to the weighted 
integrals in (\ref{rates}). We find that at order $\alpha_s$ the weight
function in the second integral is 
\begin{equation}\label{wfun}
   w(E_\gamma,E_0) = (E_\gamma-E_0) \left[ 
   1 - \frac{10}{9}\,\frac{\alpha_s(\mu)}{\pi}\,
   \ln\!\bigg( 1 - \frac{E_0}{E_\gamma} \bigg) \right] ,
\end{equation}
whereas the matching corrections are given by
\begin{eqnarray}\label{Kdef}
   K_{\rm pert}
   &=& [C_7^{(0)}(\mu)]^2\,\Bigg\{ 1
    + \frac{\alpha_s(\mu)}{2\pi}\,\Bigg[ 
    - \frac{83}{9} + \frac{4\pi^2}{9}
    + \frac{32}{3}\,\ln\frac{m_b}{\mu}
    + \frac{C_7^{(1)}(\mu)}{C_7^{(0)}(\mu)} \Bigg] \Bigg\} \nonumber\\
   &&\mbox{}+ C_2^{(0)}(\mu)\,C_7^{(0)}(\mu)\,\Bigg[
    \frac{\alpha_s(\mu)}{2\pi} \left( \mbox{Re}(r_2)
    + \frac{416}{81}\,\ln\frac{m_b}{\mu} \right)
    - \frac{\lambda_2}{9m_c^2} \Bigg] \nonumber\\
   &&\mbox{}+ C_8^{(0)}(\mu)\,C_7^{(0)}(\mu)\,
    \frac{\alpha_s(\mu)}{2\pi} \left( \frac{44}{9}
    - \frac{8\pi^2}{27} - \frac{32}{9}\,\ln\frac{m_b}{\mu} \right) ,
\end{eqnarray}
where $\mbox{Re}(r_2)\approx -4.092-12.78\,(0.29-m_c/m_b)$ \cite{GHW}.
$C_i$ are Wilson coefficients appearing in the effective weak 
Hamiltonian for $B\to X_s\gamma$ decays,\footnote{Strictly speaking, 
$C_7$ and $C_8$ are the so-called ``effective'' Wilson coefficients 
$C_7^{\rm eff}$ and $C_8^{\rm eff}$.} 
which are expanded as
\begin{equation}
   C_i(\mu) = C_i^{(0)}(\mu)
   + \frac{\alpha_s(\mu)}{4\pi}\,C_i^{(1)}(\mu) + \dots \,.
\end{equation}
Explicit expressions for these coefficients can be found, e.g., in 
\cite{CMM,Alex}. The terms in the last two lines in (\ref{Kdef}) 
result from the interference of the amplitudes corresponding to the 
operators $O_2$, $O_7$ and $O_8$ in the effective weak Hamiltonian. For 
completeness, we include a power correction proportional to 
$\lambda_2/m_c^2$ in the result for $K_{\rm pert}$, which represents a 
long-distance contribution from $(c\bar c)$ intermediate states to the 
matrix element of the operator $O_2$ \cite{Volo,Khod,Zolt,Gran,Gerh}. 
(Here $\lambda_2\approx 0.12$\,GeV$^2$ is the $B$-meson matrix element 
of the chromo-magnetic operator \cite{FaNe}.) Its numerical effect is, 
however, very small.

The above expressions for $w(E_\gamma,E_0)$ and $K_{\rm pert}$ 
are scale and scheme independent at NLO. In this context, it is crucial 
that the terms arising from operator mixing are included. The 
$\mu$-dependent terms proportional to $C_2\,C_7$ and $C_8\,C_7$ in 
(\ref{Kdef}), together with the $\mu$-dependent term in the first line, 
are required to cancel the scale dependence of the leading-order 
coefficient $[C_7^{(0)}(\mu)]^2$. Likewise, the constant terms 
proportional to $C_2\,C_7$ and $C_8\,C_7$, together with the constant 
term in the first line, are necessary to compensate the scheme 
dependence of the NLO coefficient $C_7^{(1)}$. In all previous analyses 
of the ratio (\ref{nice}), the terms proportional to $C_2\,C_7$ and 
$C_8\,C_7$ were neglected. In \cite{me}, only partial $O(\alpha_s)$ 
corrections to the $B\to X_s\gamma$ decay rate available at that time 
were included. (In addition, there is a sign mistake in the expression 
for $\eta_{\rm QCD}$ in eq.~(58) of that paper.) The NLO coefficient 
$C_7^{(1)}$ was calculated later in \cite{CMM}. In \cite{Ira}, only the 
terms in the first line in (\ref{Kdef}) were included.

In order to illustrate the importance of the various terms in 
(\ref{Kdef}), we quote the result for $\mu=m_b=4.8$\,GeV and 
$m_c/m_b=0.29$:
\begin{equation}
   K_{\rm pert}\simeq [C_7^{(0)}(m_b)]^2\,\Big(
   1 - 0.23 + 0.53\,[\mbox{2-7 mix}] + 0.03\,[\mbox{8-7 mix}] \Big)
   \approx 1.33\,[C_7^{(0)}(m_b)]^2 \,.
\end{equation}
Here $-0.23$ is the sum of the various constants in the square brackets 
in the first line in (\ref{Kdef}), and the last two terms are the 
contributions from operator mixing as shown in the second and third 
line. It is evident that the mixing contributions are numerically 
important and of opposite sign than the remaining $O(\alpha_s)$ 
corrections. Most important is the $O_2$-$O_7$ interference term, which 
is enhanced by the large ratio of Wilson coefficients 
$C_2^{(0)}(m_b)/C_7^{(0)}(m_b)\approx-3.6$.

It has recently been argued that the $B\to X_s\gamma$ decay rate should 
not be evaluated as a function of the ratio of pole masses 
$m_c/m_b\approx 0.29$, but that a more appropriate choice would be to 
use a running charm-quark mass such that $m_c/m_b\approx 0.22$ 
\cite{Misi}. If this is done, the contribution from the $O_2$-$O_7$ 
interference term in the above example increases from 0.53 to 0.65. 
Since the question of the quark-mass definition can, strictly speaking, 
only be settled by a NNLO calculation, we will include the variation of 
the result under the variation of $m_c/m_b$ between 0.22 and 0.29 as 
part of the theoretical uncertainty.

\begin{figure}
\centerline{\epsfxsize=8cm\epsffile{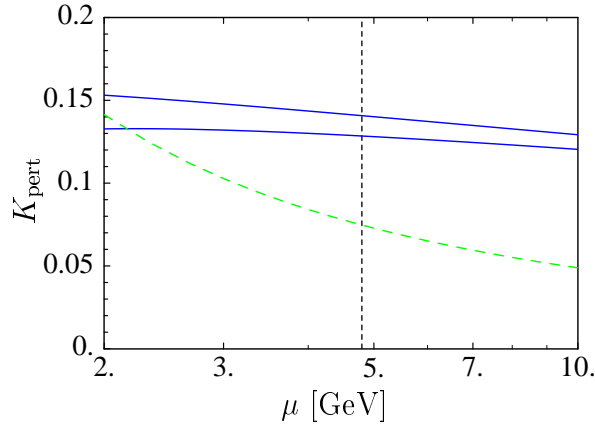}}
\centerline{\parbox{14cm}{\caption{\label{fig:Kpert}
Scale dependence of the coefficient $K_{\rm pert}$ for $m_c/m_b=0.29$ 
(lower curve) and 0.22 (upper curve). The dashed line shows the result 
without operator mixing.}}}
\end{figure}

The two solid lines in Figure~\ref{fig:Kpert} show our result for 
$K_{\rm pert}$ as a function of the renormalization scale, varied 
between 2 and 10\,GeV. The dashed line shows, for comparison, the result 
obtained ignoring the effects of operator mixing, i.e., retaining only 
the first line in (\ref{Kdef}). The vertical dashed line indicates the 
default value $\mu=m_b$ used in previous analyses. We observe a good 
stability of the NLO prediction under variation of the renormalization 
scale. The sensitivity to the charm-quark mass implies an uncertainty of 
about $\pm 5\%$. On the other hand, the result without operator mixing 
is strongly scale dependent and, for $\mu\approx m_b$, underestimates 
the correct answer by almost a factor 2.

Our result for the perturbative correction factor $K_{\rm pert}$ can be
summarized as
\begin{equation}
   K_{\rm pert} = 0.134_{\,-0.009}^{\,+0.007}\,\mbox{[scale]}
   {}_{\,-0.006}^{\,+0.007}\,\mbox{[$m_c$]}
   \pm 0.010\,\mbox{[$\alpha_s^2$]} \,.
\end{equation}
The quoted errors are obtained by scanning the renormalization scale 
$\mu$ between $m_b/2$ and $2m_b$, and the mass ratio $m_c/m_b$ between 
0.22 and 0.29. In addition, there is an uncertainty due to the neglect 
of $O(\alpha_s^2)$ corrections, which we have estimated by squaring the 
$O(\alpha_s)$ coefficients of the different combinations of Wilson 
coefficients in (\ref{Kdef}).

\begin{figure}
\centerline{\epsfxsize=8cm\epsffile{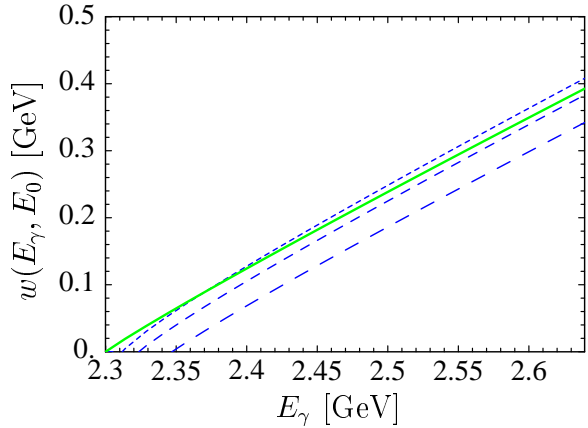}}
\centerline{\parbox{14cm}{\caption{\label{fig:resum}
Comparison of the resummed weight function in (\protect\ref{wIra}) 
(dashed lines) with the one-loop function in (\protect\ref{wfun}) 
(solid line) for $E_0=2.3$\,GeV. The singularity of the integrand is 
avoided by replacing the upper limit of integration by $0.98\,E_\gamma$ 
(long dashed), $0.99\,E_\gamma$ (dashed), and $0.995\,E_\gamma$ (short 
dashed).}}}
\end{figure}

We finally like to comment on the effect of the resummation of Sudakov
logarithms for the weight function $w(E_\gamma,E_0)$ in (\ref{wfun}). It 
is well-known that the leading, double-logarithmic contributions cancel 
in the ratio of the two rates in (\ref{nice}), so only subleading 
logarithms remain \cite{me}. These terms were resummed in \cite{Korc,Ira}
to all orders of perturbation theory. The net effect is to replace the 
weight function in (\ref{wfun}), evaluated at the scale $\mu=m_b$, by 
the more complicated function
\begin{equation}\label{wIra}
   w_{\rm res}(E_\gamma,E_0) = Z^{-1}\,
   \int\limits_{E_0}^{E_\gamma}\,dE\,
   K\bigg[ \frac{2E}{M_B} \,;\, \frac{16}{3\beta_0}\,
    \ln\bigg( 1 + \frac{\beta_0\alpha_s(m_b)}{4\pi}\,
    \ln\ln\frac{E_\gamma}{E} \bigg) \bigg] \,,
\end{equation}
where $\beta_0=11-\frac23\,n_f$. (We omit a factor 
$(1+3\bar\Lambda/M_B)\,(2 E_\gamma/M_B)^2=1+O(\Lambda_{\rm QCD}/M_B)$ 
introduced by hand in \cite{Ira}, since it is beyond the accuracy of 
(\ref{nice}) and has nothing to do with resummation effects.) The 
function $K(x;y)$ is given in eq.~(42) of \cite{Ira}, and
\begin{equation}
   Z =  1 - \frac{\alpha_s(m_b)}{2\pi} \left(
   - \frac{83}{9} + \frac{4\pi^2}{9} \right)
\end{equation}
is defined such that $w(E_\gamma,E_0)$ coincides with the one-loop 
result in (\ref{wfun}) up to terms of order $\alpha_s^2$. The integrand 
in (\ref{wIra}) is singular at $E/E_\gamma\approx 0.999$, and it was 
suggested in \cite{Ira} to avoid this singularity by replacing the 
upper limit of integration by $0.99\,E_\gamma$. It turns out that the 
numerical results are rather sensitive to this treatment. In 
Figure~\ref{fig:resum}, we compare the resummed weight function 
(\ref{wIra}) obtained with three different cutoff prescriptions with the 
one-loop function in (\ref{wfun}). We use $E_0=2.3$\,GeV corresponding
to the energy cutoff employed in the experimental analysis \cite{Ed}. 
It is evident that within the intrinsic uncertainty of the resummation 
procedure (as reflected by the sensitivity to the integration cutoff) 
the effect of Sudakov resummation is negligible. It is thus a safe 
approximation for all practical purposes to work with the fixed-order 
expression (\ref{wfun}).

In summary, we have computed the complete NLO perturbative corrections 
to the ratio of weighted integrals in (\ref{nice}). This provides the
basis for a model-independent determination of $|V_{ub}|$ from 
semileptonic and radiative $B$ decays. We find that NLO corrections 
from operator mixing are numerically important. Neglecting these terms 
would not only lead to scale and scheme-dependent predictions, but 
also introduce a numerical error in the result for $|V_{ub}|$ of as much 
as 50\% (for $\mu=m_b$). We have also shown that the resummation of 
subleading Sudakov logarithms for the weight function $w(E_\gamma,E_0)$ 
does not have a numerically significant effect on the result, although 
it is a conceptual improvement of the calculation. 

Adding the various contributions to the error in quadrature, we find 
$K_{\rm pert}=0.134\pm 0.014$ for the perturbative correction in 
(\ref{nice}). The potentially most important source of theoretical 
uncertainty is not the small perturbative error found here, but the 
presence of unknown first-order power corrections 
$\sim\Lambda_{\rm QCD}/M_B$. In practice, the empirical finding that 
the result for $|V_{ub}|$ obtained from (\ref{nice}) were independent 
of the threshold $E_0$ employed in the analysis of the experimental data 
would give us confidence that the impact of power corrections was not 
very significant.

\vspace{0.2cm}\noindent 
{\it Acknowledgments:\/}
I am grateful to Adam Leibovich, Ira Rothstein and Ed Thorndike for 
useful discussions. This work was supported in part by the National 
Science Foundation.

\vspace{0.3cm}\noindent
{\it Note added:\/}
After this note was submitted the paper \cite{Iranote} appeared, in 
which the authors suggest to raise the cutoff in the evaluation of 
(\ref{wIra}) from $0.99\,E_0$ to $0.9987\,E_0$, thereby enhancing the 
numerical effect of Sudakov resummation. Leaving aside the fact that 
this is an ad hoc prescription, we would not trust ``perturbative'' 
resummation effects that result from the immediate vicinity of the 
Landau pole in a running coupling.

\end{document}